\documentclass[twocolumn,amsmath,amssymb,floatfix,superscriptaddress,prl]{revtex4-2}

\usepackage{bm,graphicx,url,epsf,color}
\usepackage{amssymb}
\usepackage{amsthm}
\usepackage{cases}
\usepackage[colorlinks=true,linkcolor=blue,citecolor=blue]{hyperref}
\usepackage{comment}

\definecolor{purple}{rgb}{0.5,0,0.6}
\usepackage{ulem}
\definecolor{darkblue}{rgb}{0,0,0.5}
\definecolor{darkgreen}{rgb}{0,0.5,0}
\definecolor{darkred}{rgb}{.7,0,0}
\definecolor{purple}{rgb}{0.5,0,0.6}
\definecolor{orange}{rgb}{1,0.5,0}
\definecolor{grey}{rgb}{.6,.6,.6}
\definecolor{lightpink}{rgb}{1,0.7,0.75}
\definecolor{pink}{rgb}{1,0.4,0.58}
\definecolor{deeppink}{rgb}{1,0.08,0.58}

\renewcommand{\emph}[1]{\textit{#1}}

\usepackage{bbold}

\begin{document}
\title{$\mathbb{Z}_3$ parafermion in the double charge-Kondo model}


\author{D. B. Karki}
\thanks{Present address: Materials Science Division, Argonne National Laboratory, Argonne, Illinois 60439, USA}
\affiliation{Division of Quantum State of Matter, Beijing Academy of Quantum Information Sciences, Beijing 100193, China}
\author{Edouard Boulat}
\affiliation{Universit\'e  de  Paris, CNRS, Laboratoire  Mat\'eriaux  et  Ph\'enom\`enes  Quantiques, 75013  Paris,  France}
\author{Winston Pouse}
\affiliation{Department of Applied Physics, Stanford University, Stanford, CA 94305, USA}
\affiliation{SLAC National Accelerator Laboratory, Menlo Park, California 94025, USA}
\author{David Goldhaber-Gordon}
\affiliation{SLAC National Accelerator Laboratory, Menlo Park, California 94025, USA}
\affiliation{Department of Physics, Stanford University, Stanford, CA 94305, USA}
\author{Andrew K. Mitchell}\email[]{Andrew.Mitchell@UCD.ie}
\affiliation{School of Physics, University College Dublin, Belfield, Dublin 4, Ireland}
\affiliation{Centre for Quantum Engineering, Science, and Technology, University College Dublin, Ireland}
\author{Christophe Mora}\email[]{christophe.mora@u-paris.fr}
\affiliation{Universit\'e Paris Cit\'e, CNRS,  Laboratoire  Mat\'eriaux  et  Ph\'enom\`enes  Quantiques, 75013  Paris,  France}


\begin{abstract}
\noindent Quantum impurity models with frustrated Kondo interactions can support quantum critical points with fractionalized excitations. 
Recent experiments [W. Pouse et al., Nat. Phys. (2023)] on a circuit containing two coupled metal-semiconductor islands exhibit transport signatures of such a critical point. 
Here we show using bosonization that the double charge-Kondo model describing the device can be mapped in the Toulouse limit to a sine-Gordon model. 
Its Bethe-ansatz solution shows that a $\mathbb{Z}_3$ parafermion emerges at the critical point, characterized by a fractional $\tfrac{1}{2}\ln(3)$ residual entropy, and scattering fractional charges $e/3$. We also present full numerical renormalization group calculations for the model and show that the predicted behavior of conductance is consistent with experimental results.
\end{abstract}
\maketitle


Quantum impurity models, which feature a few localized, interacting quantum degrees of freedom coupled to non-interacting conduction electrons, constitute an important paradigm in the theory of strongly correlated electron systems \cite{hewson}. They describe magnetic impurities embedded in metals or other materials \cite{Kondo,costi2009kondo}, and nanoelectronic devices such as semiconductor quantum dots \cite{Goldhaber_nat(391)_1998,Cronenwett_1998,van2000kondo} or single-molecule transistors \cite{liang2002kondo,yu2005kondo}. They are also central to the understanding of bulk correlated materials through dynamical mean field theory \cite{georges1996dynamical}. Generalized quantum impurity models host a rich range of complex physics, including various Kondo effects \cite{Nozieres_Blandin_JPhys_1980,Cox_Adv_Phys(47)_1998,choi20054,paaske2006non,roch2009observation,beri2012,galpin2014conductance,keller2014emergent,mitchell2017kondo,mitchell2021so} and quantum phase transitions \cite{vojta2006impurity,jones1988low,Affleck_1992,affleck1993exact,Yuval_2003,Potok_NAT(446)_2007,mitchell2009quantum,Keller_2015,Pierre_2015,Pierre_2018}. 
Such models provide a simple platform to study nontrivial physics which can be difficult to identify in far more complex bulk materials. Indeed, exact analytical and numerical methods for quantum impurity models have given deep insights into strong correlations at the nanoscale~\cite{AFFLECK_AP_1995,andrei1984solution,EK1,wilson1975renormalization,bulla2008numerical}.

The two-channel Kondo (2CK) \cite{Nozieres_Blandin_JPhys_1980,affleck1993exact} and two-impurity Kondo (2IK) \cite{jones1988low,Affleck_1992} models are classic examples in which frustrated interactions give rise to non-Fermi liquid (NFL) physics at quantum critical points (QCPs) with fractionalized excitations. 
The seminal work of Emery and Kivelson (EK) \cite{EK1} solved the 2CK model in the Toulouse limit using bosonization techniques, and understood the QCP in terms of a free Majorana fermion localized on the impurity. 
In the 2IK model~\cite{gan1995mapping,Affleck_1992,Sela_2011,Mitchell_2012a,Mitchell_2012b,bayat2014}, a free Majorana arises from the competition between an RKKY exchange interaction coupling the impurities, and individual impurity-lead Kondo effects.
In both cases the QCP is characterized by a finite, fractional residual impurity entropy of $\tfrac{1}{2}\ln(2)$ \cite{andrei1984solution,Affleck_1992}, which is a distinctive fingerprint of the free Majorana.

Semiconductor quantum devices~\cite{Goldhaber_nat(391)_1998,Cronenwett_1998,van2000kondo} can constitute experimental quantum simulators for such impurity models, with \emph{in situ} control over parameters allowing correlated electron phenomena to be probed with precision. 
The distinctive conductance signatures predicted \cite{Yuval_2003,Sela_2011,Mitchell_2012a} for the 2CK model at criticality were since observed \cite{Potok_NAT(446)_2007,Keller_2015} (although the 2IK model has never been realized \cite{jayatilaka2011two}). More recently, Matveev's charge-Kondo paradigm \cite{Matveev_1991,Matveev1995} has emerged as a viable alternative to engineer exotic states, with both 2CK \cite{Pierre_2015} and its three-channel variant \cite{Pierre_2018} being realized experimentally.

\begin{figure}[b!]
\includegraphics[width=7.5cm]{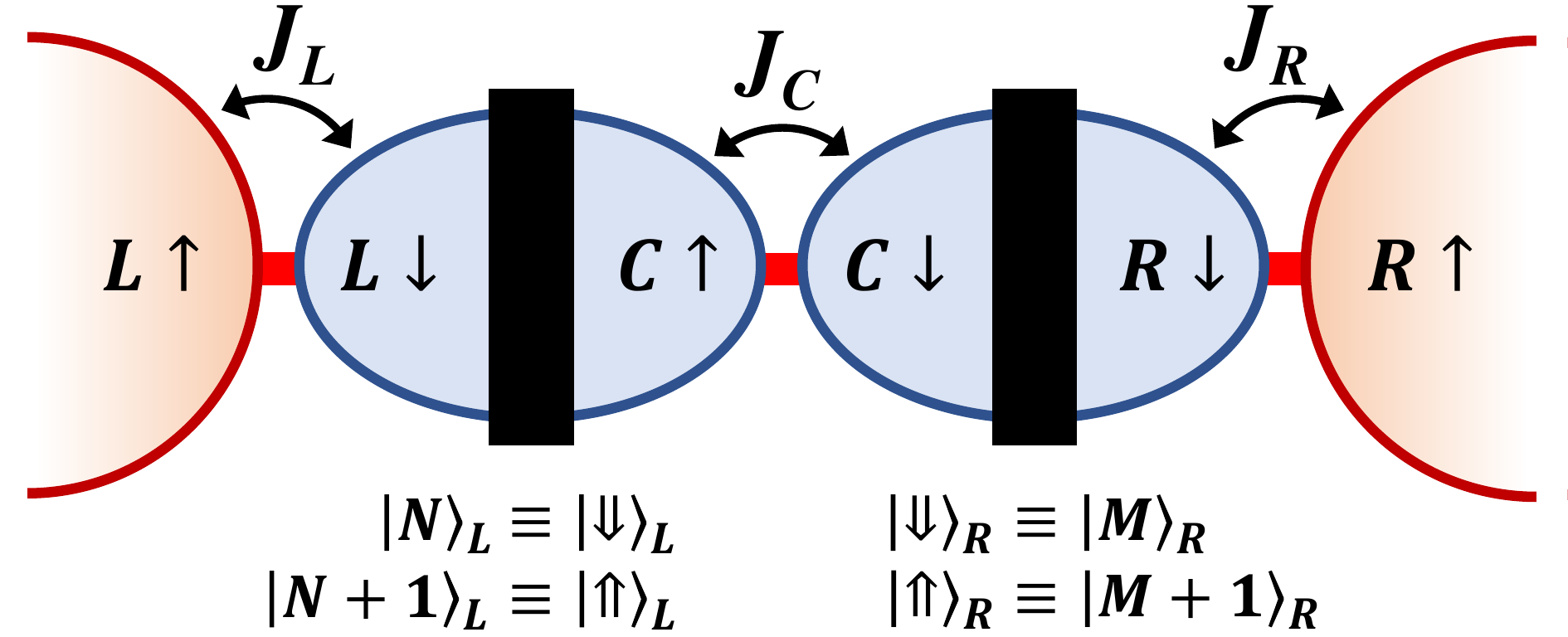}
  \caption{Schematic of the two-site charge-Kondo circuit described by the DCK model. Two hybrid metal-semiconductor islands are coupled to each other and to their own lead at QPCs. Macroscopic island charge states mapped to pseudospin degrees of freedom are flipped by tunneling at QPCs.
  }
  \label{fig:schematic}
\end{figure}

Given the intense experimental efforts to demonstrate the existence of Majoranas in quantum devices \cite{mourik2012signatures,nadj2014observation}, and the broader interest in realizing anyons for the purposes of quantum computing \cite{mzm2,hutter2016}, the Kondo route to fractionalization has gained traction \cite{lopes2020anyons,komijani2020isolating,gabay2022multi,lotem2022}. Experimental circuit realizations of more complex quantum impurity models offer the tantalizing opportunity to produce more exotic anyons in tunable nanoelectronics devices. 
This can be viewed as part of a wider effort to study fractionalization in condensed matter systems \cite{wakatsuki2014,tsvelik2022,coleman2022,delpozo2022,klinovaja2014,cheng2014,mazza2018}.

However, despite the suggestive fractional entropies in certain Kondo-type models \cite{andrei1984solution,AFFLECK_AP_1995,beri2017,altland2014bethe,Pierre_2018,pouse2023,Note1}, the \textit{explicit} construction of parafermion operators in these systems has not previously been possible. This is because -- unlike for the simpler case of Majoranas -- parafermions cannot arise in an effective free fermion system. Applying the EK method yields an irreducibly strongly-interacting model, which has hitherto hindered finding exact solutions in which free local parafermions could be identified.

In this Letter we study the double charge-Kondo (DCK) model describing a very recent experiment~\cite{pouse2023} involving two hybrid metal-semiconductor islands coupled together in series, and each coupled to its own lead, at quantum point contacts (QPCs) -- see Fig.~\ref{fig:schematic}. The DCK model is a variant of the celebrated 2IK model, but with an inter-island Kondo interaction
rather than an RKKY exchange interaction~\cite{jones1988low}.
At the triple point in the charge stability diagram of the device, a QCP was found to arise due to the competition between island-lead Kondo and inter-island Kondo~\cite{pouse2023}. Numerical renormalization group \cite{wilson1975renormalization,bulla2008numerical,mitchell2014generalized,stadler2016interleaved} (NRG) calculations for the DCK model showed a fractional residual entropy of $\tfrac{1}{2}\ln(3)$ at the QCP -- suggesting an unusual anyonic state (and not simply a Majorana). 
The same critical point and fractional entropy were identified analytically near perfect QPC transmission \cite{KBM}, although no Kondo effects occur in this limit.

Here we examine the ``Kondo'' case of weak-to-intermediate transmission, and apply the EK mapping \cite{EK1} in the Toulouse limit. Even though the EK method yields a highly nontrivial interacting model, we show that it can nevertheless be solved using Bethe ansatz. Instead of the free Majorana found by EK for the 2CK model, we explicitly establish the existence of a $\mathbb{Z}_3$ parafermion in the DCK model, and identify it as the source of the $\tfrac{1}{2}\ln(3)$ residual entropy. Analytic expressions for conductance near the QCP are also extracted, and we show that experimental transport data are consistent with these predictions. To complete the theoretical description, we obtain the full temperature dependence of entropy and conductance via NRG, which does not rely on the Toulouse approximation. 


\emph{System and model.--}
The two-island circuit illustrated in Fig.~\ref{fig:schematic} is described by the DCK model at low temperatures $T\ll E_C$ (with $E_C$ the island charging energies) for weak-to-intermediate QPC transmissions, see Ref.~\cite{pouse2023}:
\begin{eqnarray}\label{eq:Hdck}
\begin{split}
H_{\rm DCK}=&\Big(J_{\rm L}^{\phantom{+}} S_{\rm L}^+s_{\rm L}^-+J_{\rm R}^{\phantom{+}} S_{\rm R}^+s_{\rm R}^-+J_{\rm C}^{\phantom{+}} S_{\rm R}^+ S_{\rm L}^-s_{\rm C}^-+{\rm H.c.}\Big)~~~ \\
& - h_L^{\phantom{z}} S_L^z - h_R^{\phantom{z}} S_R^z + I S_L^z S_R^z + H_{\rm elec} \;,
\end{split}
\end{eqnarray}
where $H_{\rm elec}=\sum_{\alpha,\sigma,k} \epsilon_k^{\phantom{\dagger}} \psi_{\alpha\sigma k}^{\dagger}\psi_{\alpha\sigma k}^{\phantom{\dagger}}$ describes the electronic reservoirs either side of QPC $\alpha=L,C,R$. Although the physical electrons are spin-polarized~\cite{pouse2023}, we label electrons on the lead or island either side of QPC $L,R$ as $\sigma=\uparrow$ or $\downarrow$, and island electrons to the left or right of the central QPC $C$ as $\sigma=\uparrow$ or $\downarrow$ -- see Fig.~\ref{fig:schematic}. We assume linear dispersion $\epsilon_k = v_F k$, with momentum $k$. We then define pseudospin operators $s_{\alpha}^-=\psi_{\alpha\downarrow}^{\dagger}(0)\psi_{\alpha\uparrow}^{\phantom{\dagger}}(0)$ and $s_{\alpha}^+=(s_{\alpha}^-)^{\dagger}$, where $\psi_{\alpha\sigma}(0)$ is defined at the QPC position. 
Confining our attention to the lowest two macroscopic charge states of each island $|n,m\rangle\equiv |n\rangle_L\otimes |m\rangle_R$, with $n=N,N+1$ the number of electrons on the left island and $m=M,M+1$ electrons on the right island, we introduce `impurity' charge pseudospin operators $S_L^+=\sum_m |N+1,m\rangle\langle N,m|$, $S_L^z=\sum_m \tfrac{1}{2}[|N+1,m\rangle\langle N+1,m| - |N,m\rangle\langle N,m|]$, $S_R^+=\sum_n |n,M+1\rangle \langle n,M|$, $S_R^z=\sum_n \tfrac{1}{2}[|n,M+1\rangle\langle n,M+1| - |n,M\rangle\langle n,M|]$ and $S_{\alpha}^-=(S_{\alpha}^+)^{\dagger}$. 
The first line in Eq.~\eqref{eq:Hdck} therefore corresponds to tunneling processes at the three QPCs (with the tunneling amplitude $J_{\alpha}$ being related to the transmission $\tau_{\alpha}$ of QPC $\alpha$). Gate voltages on the islands control $h_{L,R}$ and allow the charge stability diagram to be navigated. $I$ is a capacitive interaction between the two islands. For $J_{L,C,R}=I=0$, the four retained charge configurations $|n,m\rangle$ are degenerate when $h_L=h_R=0$. However, a finite $J_C$ and/or $I$ partially lifts this degeneracy to yield a pair of separated triple points (TPs) in gate voltage space. As with the experiment \cite{pouse2023}, here we focus on the vicinity of the TP at which the charge configurations $|N,M\rangle/|N+1,M\rangle/|N,M+1\rangle$ are degenerate. We hereafter neglect the term $I$, since it just renormalizes the TP splitting already induced by $J_C>0$ and is otherwise irrelevant \cite{KBM}. The rest of this Letter is devoted to the nontrivial Kondo competition arising when the couplings to the leads are switched on, $J_{L,R} >0$.


\emph{QCP.--} At the TP, the three `impurity' states (the degenerate charge configurations of the two-island structure) are interconverted by tunneling at the three QPCs:
\\ $~~~~~~~~|N,M\rangle \overset{J_L}{\leftrightarrow} |N+1,M\rangle \overset{J_C}{\leftrightarrow} |N,M+1\rangle \overset{J_R}{\leftrightarrow} |N,M\rangle$\\ 
The accompanying conduction electron pseudospin-flip scattering at each QPC described by the operators $s_{L,C,R}^{\pm}$ in Eq.~\eqref{eq:Hdck} give rise to competing Kondo effects. Since island-lead and inter-island Kondo effects cannot be simultaneously satisfied, a frustration-driven QCP arises when $J_L=J_R=J_C$, as reported in Refs.~\cite{pouse2023,KBM}.


\begin{figure*}[ht!]
\includegraphics[width=18cm]{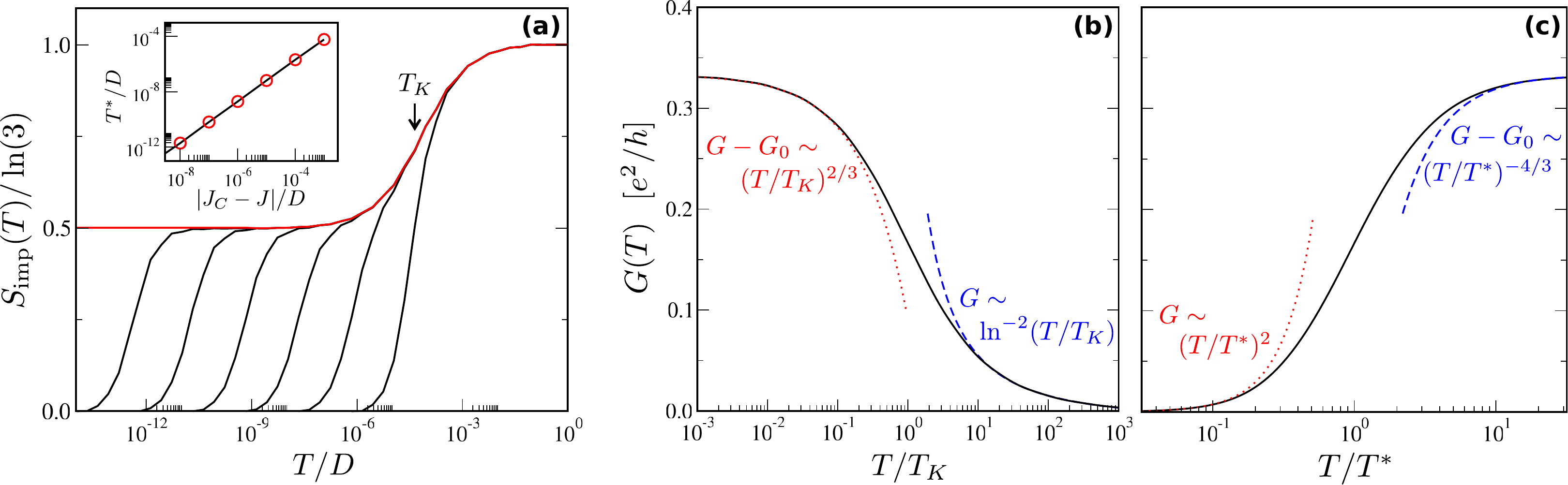}
  \caption{NRG results at the triple point of the DCK model. (a) Entropy $S_{\rm imp}(T)$ in the vicinity of the critical point, showing the flow $\ln(3) \to \tfrac{1}{2}\ln(3)$ on the Kondo scale $T_K$, and subsequently $\tfrac{1}{2}\ln(3) \to 0$ on the Fermi liquid scale $T^*$. Plotted for $J/D=0.2$ and $|J_C-J|/D=10^{-3}, 10^{-4}, ..., 10^{-8}$ (black lines) approaching the critical point $J_C=J$ (red line). $D$ is the conduction electron bandwidth. Inset shows the power-law behavior Eq.~\eqref{eq:Tstar}. (b) Universal conductance curve as a function of $T/T_K$ at the critical point; (c) Universal Fermi liquid crossover as a function of $T/T^*$. Conductance asymptotes are discussed in the text.
  }
  \label{fig:nrg}
\end{figure*}

\emph{NRG solution.--} In Fig.~\ref{fig:nrg} we present numerically-exact results for the DCK model tuned to the TP, obtained by NRG \cite{wilson1975renormalization,bulla2008numerical,mitchell2014generalized,stadler2016interleaved} (see \footnote{See Supplemental Material for details with Refs.~\cite{FG1,Furusaki1995b,morel2021,Sela_2018,weichselbaum2007sum,Sela_2016,Galpin2014,anders2006spin,transport,affleck2001} }
for details). 
We set $J_L=J_R\equiv J$ and vary $J_C$ in the vicinity of the QCP arising when $J_C=J$. In panel (a) we show the impurity contribution to the entropy $S_{\rm imp}$ as a function of temperature $T$. The critical point $J_C=J$, shown as the red line, exhibits Kondo `overscreening' to an NFL state on the scale of $T_K$. The three degenerate charge states give a high-$T$ entropy of $\ln(3)$, but the entropy is partially quenched to $\tfrac{1}{2}\ln(3)$ for $T\ll T_K$. Introducing channel anisotropy $J_C \ne J$ induces a Fermi liquid (FL) crossover on the lower scale of $T^*$, below which the entropy is completely quenched. The inset shows the extracted power-law behavior,
\begin{eqnarray}\label{eq:Tstar}
T^*/T_K \sim (|J_C-J|/T_K)^{3/2} \;.
\end{eqnarray}
The same form was reported for detuning away from the TP in Ref.~\cite{pouse2023}. For $|J_C-J| \ll T_K$ we have good scale separation $T^*\ll T_K$, such that the crossover to the critical point is a universal function of the single scaling parameter $T/T_K$, whereas the crossover away from it is a universal function of only $T/T^*$. This is reflected in the behavior of series conductance, shown in panels (b,c). At the highest temperatures $T\gg T_K$, Kondo-renormalized spin-flip scattering gives standard $\ln^{-2}(T/T_K)$ corrections to conductance; whereas on the lowest temperature scales $T\ll T^*$, we observe conventional FL scaling of conductance $\propto (T/T^*)^2$. Much more interesting is the behavior in the vicinity of the critical fixed point \cite{pouse2023},
\begin{subnumcases}
{G_0-G(T) \sim \label{eq:G}}
  (T/T_K)^{2/3}\;,   & ~~$T \ll T_K$ \label{eq:GTk}  
 \\
  (T/T^*)^{-4/3}\;,  & ~~$T \gg T^*$ \label{eq:GTstar}
\end{subnumcases}
with $G_0=e^2/3h$. Eqs.~\eqref{eq:Tstar}, \eqref{eq:G} are also  obtained analytically and discussed in the following.


\emph{Bosonization and Toulouse point.--}
We now turn to the details of our exact solution. Following  EK~\cite{EK1}, we bosonize the conduction electron Hamiltonian $H_{\rm elec}$ and obtain a simplified model in the Toulouse limit after applying a unitary transformation. 

As a first step we write $\psi_{\alpha\sigma}=e^{i\phi_{\alpha\sigma}}/\sqrt{a}$ with $a= 4\pi v_{\rm F}\equiv 1$ and introduce three chiral bosonic fields $\delta\phi_{\alpha}\equiv \left(\phi_{\alpha \uparrow}-\phi_{\alpha \downarrow}\right)/\sqrt2$ for $\alpha=L,R,C$. The conduction electron pseudospin operators follow as $s_{\alpha}^-=e^{ i\sqrt2 \delta\phi_{\alpha}}$, and 
\begin{equation}
H_{\rm elec}=\frac{v_{\rm F}}{4\pi}\sum_\alpha\int dx\;\left(\frac{\partial \delta\phi_\alpha}{\partial x}\right)^2.
\end{equation}
For $h_L=h_R=I=0$, we can cast the DCK model as,
\begin{align}
H_{\rm DCK}=H_{\rm elec}+\Big[ & J_{\rm L}^{\phantom{+}} S_{\rm L}^+e^{i\sqrt2 \delta\phi_{\rm L}}+J_{\rm R}^{\phantom{+}} S_{\rm R}^+e^{i\sqrt2 \delta\phi_{\rm R}}\nonumber\\
& +J_{\rm C}^{\phantom{+}} S_{\rm R}^+ S_{\rm L}^-e^{i\sqrt2 \delta\phi_{\rm C}}+{\rm H.c.}\Big],
\end{align}
where all fields are implicitly taken at $x=0$. To make progress, we deform the original DCK model, which features only transverse couplings $J_{\alpha}$, by adding an Ising term $\bar{H}_{\rm DCK} = H_{\rm DCK} + H_{\rm I}$. Since pseudospin anisotropy is RG irrelevant, $H_{\rm I}$ affects only the flow, not the stable fixed point itself. Therefore the critical fixed point (and indeed the entire FL crossover in the limit $T^*\ll T_K$ \cite{Sela_2011,Mitchell_2012a}) is the same for any choice of $H_{\rm I}$. We shall exploit this property to identify an exactly-solvable Toulouse limit.
To do this we effect a change of basis,
\begin{align}
\delta\phi_{\rm A} & =\left(\delta\phi_{\rm R}-\delta\phi_{\rm C}-\delta\phi_{\rm L}\right)/\sqrt3,\nonumber\\
\delta\phi_{\rm B} & =\left(\delta\phi_{\rm L}+\delta\phi_{\rm R}\right)/\sqrt2,\nonumber\\
\delta\phi_{\rm D} & =\left(\delta\phi_{\rm L}-2\delta\phi_{\rm C}-\delta\phi_{\rm R}\right)/\sqrt6\label{bfields}
\end{align}
and introduce 
$\delta \phi_{1/2} = \frac{\delta \phi_{\rm B}}{\sqrt{2}} \pm \frac{\delta \phi_{\rm D}}{\sqrt{6}}$. We now choose,
\begin{equation}
    H_{\rm I} = \lambda \left [S^z_{\rm L}\partial_x \delta\phi_1(0)+ S^z_{\rm R}\partial_x\delta\phi_2 (0) \right ],
\end{equation}
and rotate the Hamiltonian into $U \bar{H}_{\rm DCK} U^\dagger = H_{\rm elec} + H_{\rm EK}$ using the EK 
unitary transformation~\cite{EK1}
\begin{equation}\label{ekt}
U=\exp\left[-i \tfrac{1}{\sqrt{2}} \{S^z_{\rm L} \delta\phi_1(0)+S^z_{\rm R} \delta\phi_2(0)\}\right].
\end{equation}
We then obtain
\begin{eqnarray}\label{EKhamil}
\begin{split}
H_{\rm EK} = & \Big[J_{\rm L}^{\phantom{+}} S_{\rm L}^- + J_{\rm R}^{\phantom{+}} S_{\rm R}^+ + J_{\rm C}^{\phantom{+}} S_{\rm L}^+  S_{\rm R}^-\Big] e^{i\sqrt{\frac{2}{3}}\delta\phi_{\rm A}} +{\rm H.c.} \\[2mm]
& + \bar{\lambda} \left [S^z_{\rm L}\partial_x \delta\phi_1(0) + S^z_{\rm R} \partial_x \delta\phi_2 (0) \right ]
\end{split}    
\end{eqnarray}
where $\bar{\lambda} = \lambda - 1/(4 \pi)^2$. The Toulouse limit is obtained by setting $\bar{\lambda} =0$,
for which the bosonic modes $\delta\phi_{\rm B, \rm D}$ fully decouple and remain free. The symmetric charge mode $\delta\phi_{\rm A}$ thus controls the low-energy behavior following Kondo screening. At the QCP with isotropic  couplings $ J_L = J_R = J_C \equiv J$, the model further simplifies,
\begin{equation}\label{eq:EK}
H_{\rm EK}=J\; \sigma \, e^{i\sqrt{\frac{2}{3}}\delta\phi_{\rm A}}+{\rm H.c.},\;\;\;\;\;\sigma=\left(
\begin{array}{ccc}
 0 & 1 & 0 \\
 0 & 0 & 1 \\
 1 & 0 & 0 \\
\end{array}
\right),
\end{equation}
where the operator $\sigma$ circularly permutes the three  impurity states $|N,M\rangle/|N+1,M\rangle/|N,M+1\rangle$.


\emph{Parafermion modes.--} In analogy with the description of chiral Potts (clock) models by parafermionic chains~\cite{Fendley2012}, we define a second operator $\tau = {\rm diag} (1,\omega,\omega^2)$, with $\omega = e^{2 i \pi/3}$ in the impurity subspace. The  operators~\cite{Fendley2012,alicea2016} $\sigma$ and $\sigma' = \sigma \tau$ then obey the parafermionic properties, 
\begin{equation}
\sigma^3=\sigma'^3=1, \qquad \sigma \sigma' = \omega \sigma' \sigma,
\end{equation}
and thereby generalize the Majorana operators to a $3$-dimensional space with circular $\mathbb{Z}_3$ symmetry. 

Importantly, $H_{\rm EK}$ includes only the terms $\sigma$ and $\sigma^\dagger$, and not $\sigma'$ (Eq.~\eqref{eq:EK}). Since $\sigma \sigma^\dagger = \sigma^\dagger \sigma$, the parafermion $\sigma$ commutes with $H_{\rm EK}$ and remains \emph{free}. Conversely, $\sigma'$ does not commute and it acquires a finite scaling dimension. 


\emph{Sine-Gordon model and Bethe-ansatz solution.--}
We rotate to the simultaneous eigenbasis of $\sigma$ and $\sigma^\dagger$ and write
$H_{\rm EK}= H^0\oplus H^+\oplus H^-$, with
\begin{equation}\label{eq:hbasis}
H^r= 2J \cos\left(\sqrt{\frac{2}{3}}\delta\phi_A+r\frac{2\pi}{3}\right) \;, \qquad r=0,\pm 1\;.
\end{equation}
The DCK model reduces to three decoupled boundary sine-Gordon models \cite{Kane_1992, Chamon_1995, Fendley_PRL_1995, Fendley_PRB_1995}, related to each other by a $\mathbb{Z}_3$ circular shift of the field $\delta \phi_A \to \delta \phi_A + 2\pi/\sqrt{6}$. They all have the same Bethe-ansatz solution describing the crossover from high to low energies (the same crossover as an impurity in a one-dimensional electron gas with Luttinger parameter $K=1/3$  ~\cite{Kane_1992}). In particular, the residual entropy is predicted~\cite{fendley1994,fendley2007} to decrease by $\Delta S = \tfrac{1}{2}\ln(3)$ along the crossover. For the DCK model we therefore expect a crossover in the impurity entropy from $\ln(3)$ to $\tfrac{1}{2}\ln(3)$, as confirmed by the NRG results in Fig.~\ref{fig:nrg}(a). The parafermions $\sigma$ and $\sigma'$ generate the threefold charge subspace. Since $\sigma'$  is screened but $\sigma$ remains free, it simply halves the residual entropy. The same residual entropy was found in the quasi-ballistic limit \cite{KBM}.


\emph{Conductance at the critical point.--}
The linear conductance between left and right leads is obtained from the Kubo formula $G=-2\pi \lim_{\omega\to 0} [{\rm Im}~K(\omega)/\omega]$, with $K(\omega)$ the Fourier transform of the retarded current-current correlator $K(t)=-i \theta(t)\langle [ I(t), I(0) ]\rangle$. Following the above mapping, $I=-\tfrac{e}{2\pi}\sqrt{\tfrac{2}{3}} \partial_t \Theta_A$, where $\Theta_A$ is the field conjugate to $\delta \phi_A$. Since $\delta \phi_A$ is pinned at the critical fixed point, $\Theta_A$ is free, and so $\langle \Theta_{\rm A} (t) \Theta_{\rm A} (0) \rangle \sim - \ln t$ at $T=0$. This yields $G=G_0=e^2/3h$: out of the three fields, only $\phi_A$ appears in Eq.~\eqref{EKhamil}, thus only $\Theta_A$ transports electrons, yielding $1/3$ of a perfect conductance.


\emph{Conductance scaling in the Kondo regime, $T\ll T_{K}$.--}
We now turn to the leading finite-temperature corrections to the $T=0$ conductance at the critical point. To do this, we must perturb away from the exactly solvable EK point by reintroducing finite $\bar{\lambda}$. This is because the RG flow to the critical fixed point is affected by $\bar{\lambda}$. The leading irrelevant operator (LIO) at the QCP is given by 
$ \mathcal{O}_{\rm LIO} = \bar{\lambda} [S^z_{\rm L}
    \partial_x \delta\phi_1(0)+ S^z_{\rm R} \partial_x \delta\phi_2 (0)]$. As we show in \cite{Note1}, the operators $\partial_x \delta\phi_{1,2}(0)$ both have scaling dimension 1 and $S^z_{\rm L,R}$ have scaling dimension $\tfrac{1}{3}$. This yields $\Delta_{\rm LIO}=4/3$, and therefore allows us to identify the leading correction to conductance (arising at order  $\bar{\lambda}^2$) as $\delta G\sim \left(T/T_{\rm K}\right)^{2\left(\Delta_{\rm LIO}-1\right)}$  \cite{Note1}, which  reproduces Eq.~\eqref{eq:GTk}. 


\emph{FL crossover.--}
The QCP is destabilized by gate voltage detuning away from the TP (appearing as pseudo-Zeeman fields $h_{L,R}$ in the DCK model), or by channel anisotropy
$\delta J$. 
The resulting FL crossover is controlled by the FL scale $T^*$. Assuming $T^*\ll T_K$, we may again utilize the Toulouse limit and set $\bar{\lambda}=0$ to analyze the FL crossover, since any finite $\bar{\lambda}$ scales to zero anyway under RG for $T \ll T_K$. 
Both perturbations $h_{L,R}$ and $\delta J$ have the effect of coupling the otherwise independent sectors of $H_{\rm EK}$ given by Eq.~\eqref{eq:hbasis}. We focus here on finite $h_R$ for simplicity. From Eq.~\eqref{eq:Hdck}, $h_R$ couples to $S_{\rm R}^z$, which in the rotated basis is given by $S_{\rm R}^z=\tfrac{1}{3}(\omega \tau + \omega^* \tau^{\dagger})$. Analyzing its action at the QCP~\cite{Note1}, we identify $\tau= e^{-i\sqrt{2/3} \Theta_A}$, where this operator circularly permutes the sectors $r$ in Eq.~\eqref{eq:hbasis}. $S_{\rm R}^z$ thus inherits the RG-relevant scaling dimension $\Delta_R=1/3$ of $\tau$,  such that
finite $h_R$ generates a FL scale $T^* \sim h_{R}^{1/(1-\Delta_R)}$. Since $S_{\rm L}^z$ and $\delta J$ have the same scaling dimension $\Delta_R$, in general we have $T^* \sim (h_L^{3/2}, h_R^{3/2}, \delta J^{3/2})$
\cite{Note1}, which reduces to Eq.~\eqref{eq:Tstar} in the case of pure channel anisotropy. 
The leading correction in $T/T^*$ to the QCP conductance $G_0$ then follows as $\delta G\sim  \left(T/T^*\right)^{2\left(\Delta_{\rm R}-1\right)}$, yielding Eq.~\eqref{eq:GTstar}. Additionally, the free parafermion at the QCP is shown by noise calculation~\cite{Note1} to scatter fractional charges $e^*=e/3$.


\begin{figure}[b!]
\includegraphics[width=7.5cm]{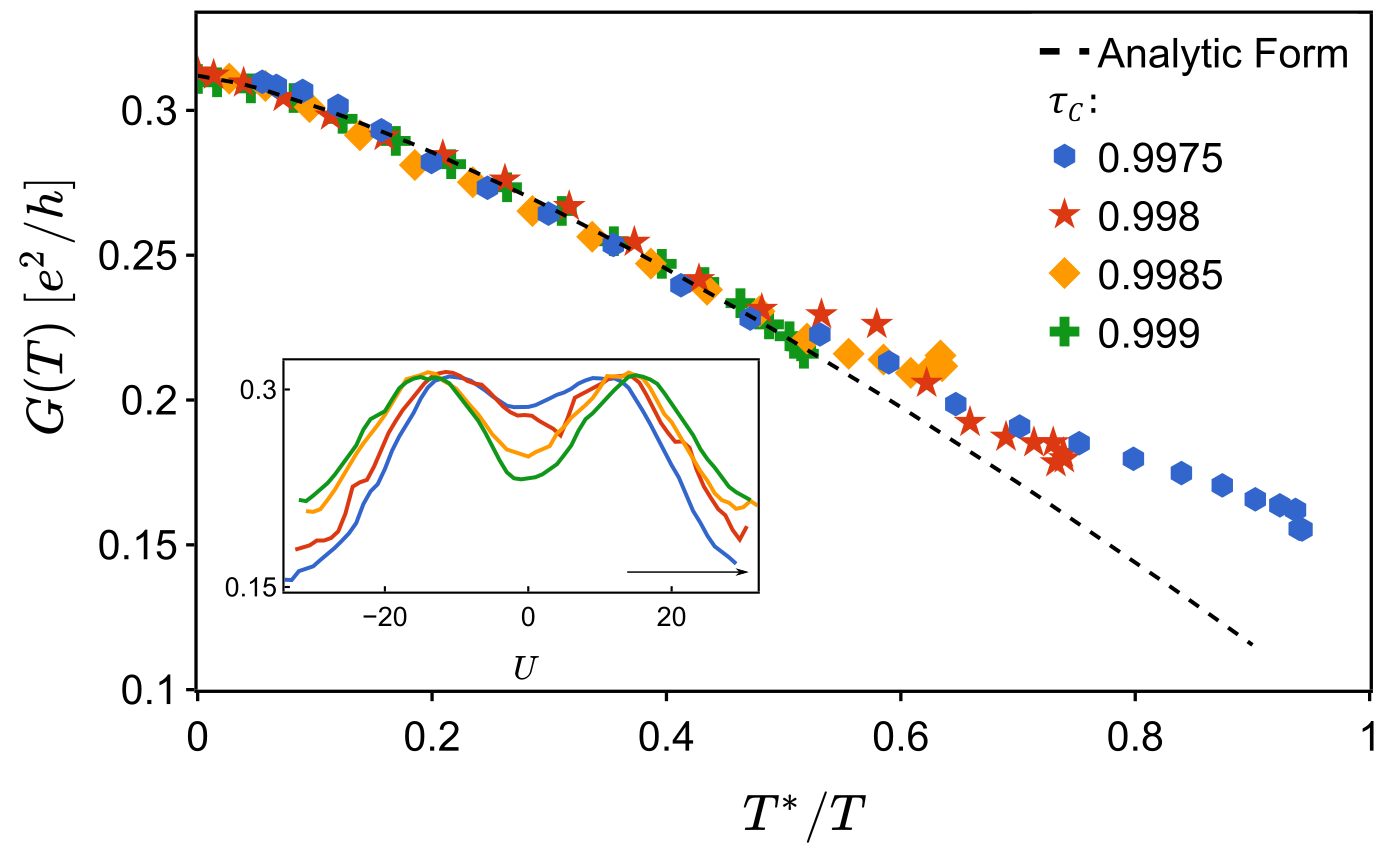}
  \caption{Scaling collapse of experimental data from Ref.~\cite{pouse2023} (points) onto the asymptotic form of Eq.~\eqref{eq:GTstTk} (dashed line). Experiment performed for $\tau=0.95$ at $T=20$mK with different $\tau_C$. Inset shows experimental data along a line cutting through a pair of triple points. The main figure data are taken from a subset of this, starting at a triple point and moving toward higher $U$, indicated by the arrow. This avoids the discontinuity near $U=0$ due to a charge switching event present in the red curve. See \cite{Note1,pouse2023} for experimental details, determination of $T^*$, and fitting procedure. }
  
  \label{fig:expt}
\end{figure}

\textit{Comparison with experiment.--} Finally we turn to the implications of our results for the experiments of Ref.~\cite{pouse2023}. Although the experimental results  were obtained at large transmission $\tau$, $\tau_C$, we expect the universal low-temperature behavior near the QCP to be the same as that discussed above for the Kondo limit \cite{KBM,Note1}. Since the maximum conductance measured is slightly lower than the predicted value $G_0=e^2/3h$, we infer that the quantum critical state is not fully developed at experimental base temperatures. Detuning away from the TP by varying the island gate voltages $U$ generates a pseudo-Zeeman field $h_L=h_R$ in the DCK model, whereas detuning QPC transmission $\tau_C$ (while keeping $\tau$ constant) maps to channel anisotropy $\delta J$. Either destabilizes the QCP and generates a finite FL scale $T^*$. Without perfect scale separation, we expect,
\begin{eqnarray}\label{eq:GTstTk}
G_0-G(T) \sim  c_K(T/T_K)^{2/3}+ c_*(T/T^*)^{-4/3} \;.
\end{eqnarray}
In Fig.~\ref{fig:expt} we plot the experimental data vs $T^*/T$, with $T^*$ estimated \cite{Note1,pouse2023} for each combination of $\tau_C$ and $U$, while $T=20$mK is kept fixed. The rescaled data compare well with Eq.~\eqref{eq:GTstTk}, when $c_K/T_K^{2/3}$ and $c_*$ are used as free fit parameters. This provides strong evidence that the vicinity of the QCP in the DCK model is probed experimentally in the device of Ref.~\cite{pouse2023}.


\textit{Conclusion and outlook.--} 
The two-site charge-Kondo setup described by the DCK model can in the Toulouse limit be mapped to a solvable boundary sine-Gordon model by bosonization methods. At the QCP we show that the residual entropy $\tfrac{1}{2}\ln(3)$ is due to a free $\mathbb{Z}_3$ parafermion, while a second parafermion mode is Kondo screened. Exploiting the mapping, we also obtain exact results for the conductance near the critical point that agree not only with NRG results but also with experimental data. This suggests that a $\mathbb{Z}_3$ parafermion is already present in the experimentally-measured device of Ref.~\cite{pouse2023}. 
This could be demonstrated more explicitly by measuring experimentally the fractional entropy of the parafermion using the methods proposed and implemented in Refs.~\cite{sela2019detecting,child2021entropy,han2022fractional}. Our approach also opens the door to studying other phases of quantum matter with irreducible strong interactions using the Emery-Kivelson mapping.

\vfill


\begin{acknowledgments}
 \textit{Acknowledgments.--} This work was supported by the Irish Research Council through the Laureate Award 2017/2018 grant IRCLA/2017/169 (AKM) and the French National Research Agency (project SIMCIRCUIT, ANR-18-CE47-0014-01). Analysis/extraction of experimental data compared to scaling theory in Fig.~\ref{fig:expt} was supported by the U.S. Department of Energy (DOE), Office of Science, Basic Energy Sciences (BES), under Contract DE-AC02-76SF00515
\end{acknowledgments}


\bibliography{biblio}

\onecolumngrid

\clearpage

\setcounter{figure}{0}
\setcounter{section}{0}
\setcounter{equation}{0}
\renewcommand{\theequation}{S\arabic{equation}}
\renewcommand{\thefigure}{S\arabic{figure}}

\onecolumngrid

\section*{\Large{Supplemental material}}

\section{Triplet point splitting}\label{secii}

The model for the double-charge Kondo experiment~\cite{pouse2023} is detailed in the main text following Eq.~(1). Its Hamiltonian takes the form
\begin{eqnarray}\label{eq:Hdck2}
H_{\rm DCK}=\Big(J_{\rm L}^{\phantom{+}} S_{\rm L}^+s_{\rm L}^-+J_{\rm R}^{\phantom{+}} S_{\rm R}^+s_{\rm R}^-+J_{\rm C}^{\phantom{+}} S_{\rm R}^+ S_{\rm L}^-s_{\rm C}^-+{\rm H.c.}\Big)  - h_L^{\phantom{z}} S_L^z - h_R^{\phantom{z}} S_R^z + I S_L^z S_R^z + H_{\rm elec} \;,
\end{eqnarray}
and couples the four charge states of the two islands 
\begin{equation}\label{eq:states}
    |N,M\rangle, \qquad |N+1,M\rangle, \qquad  |N,M+1\rangle, \qquad |N+1,M+1\rangle
\end{equation}
to electron tunneling at the three QPC with amplitudes $J_L$, $J_C$ and $J_R$. The four charge states are conveniently identified with two pseudospins onto which the pseudospin-$\tfrac{1}{2}$ operators $S_{L/R}^{\pm},S_{L/R}^z$ act. In the absence of electron tunneling, $J_{L/R/C}=0$, the four charge states are degenerate when $h_L$, $h_R$ and $I$ are all vanishing, and define a quadruplet point. However, finite tunneling amplitudes $J_{L/R/C} \ne 0$ resonantly couple the two state triplets $|N,M\rangle/|N+1,M\rangle/|N,M+1\rangle$ and $|N,M+1\rangle/|N+1,M\rangle/|N+1,M+1\rangle$. As the pair of states $|N+1,M\rangle/|N,M+1\rangle$ participates in the two triplets, their energies are decreased with respect to the states $|N,M\rangle$ and $|N+1,M+1\rangle$, effectively the same effect as the capacitive interaction $I>0$. As a result, finite $J_{L/R/C}$ and $I$ split the quadruplet into two triplets states moving symmetrically along the line $h_L=h_R$. 

The splitting is confirmed in NRG calculations. It has also been derived in the quasi-ballistic limit of very open QPC~\cite{KBM}. In the main text, we focus on the region in the stability diagram close to one triplet point and correspondingly shift the definitions of $h_L$ and $h_R$ such that this triplet point is located at the origin $h_L=h_R=0$. The capacitive interaction $I$ is moreover discarded as it simply moves the location of the triplet point.

\section{$\mathbb{Z}_3$ parafermion}\label{seciv}

Following the general idea of Emery and Kivelson~\cite{EK1}, the Toulouse limit is obtained by bosonization of the DCK model (see main text) with the addition of an irrelevant term to reach an exactly solvable line. In the DCK model, the solvable line is a boundary sine-Gordon Hamiltonian in three decoupled charge sectors with Luttinger parameter $1/3$. Remarkably, the study of the corresponding infrared fixed point and its vicinity retrieves many NRG findings and scalings, which confirms that the term added to the Hamiltonian is indeed irrelevant and does not affect the low-energy behaviour. It also identifies a fractional entropy $\frac 1 2 \ln 3$ associated with a free, or unscreened, $\mathbb{Z}_3$ parafermion. This result is completely analogous to, and in fact extends, the unscreened $\mathbb{Z}_2$ Majorana fermion emerging in the two-channel Kondo model~\cite{EK1}.

\subsection{Parafermion screening}

We introduce two parafermion (clock) operators $\sigma$ and $\sigma' = \sigma \tau$. In the original charge basis $|N,M\rangle/|N+1,M\rangle/|N,M+1\rangle$, they act as matrices
\begin{equation}
  \sigma=\left(
\begin{array}{ccc}
 0 & 1 & 0 \\
 0 & 0 & 1 \\
 1 & 0 & 0 \\
\end{array}
\right) \qquad  \qquad   \tau=\left(
\begin{array}{ccc}
 1 & 0 & 0 \\
 0 & \omega & 0 \\
 0 & 0 & \omega^2 \\
\end{array}
\right)
\qquad \qquad  \sigma'=\left(
\begin{array}{ccc}
 0 & \omega & 0 \\
 0 & 0 & \omega^2 \\
 1 & 0 & 0 \\
\end{array}
\right) 
\end{equation}
with $\omega =e^{2 i \pi/3}$, and obey the parafermionic rules 
\begin{equation}
\sigma^3=\sigma'^3=1, \qquad \sigma \sigma' = \omega \sigma' \sigma.
\end{equation}
They generalize the Majorana operators to the $3$-dimensional space compatible with the circular $\mathbb{Z}_3$ symmetry. $\sigma$ and $\sigma'$ generate the charge subspace in the sense that this three-dimensional subspace has the minimal dimension to support the above parafermionic properties.

The Hamiltonian on the Toulouse line $H_{\rm EK} = J\; \sigma \, e^{i\sqrt{\frac{2}{3}}\delta\phi_{\rm A}}+{\rm H.c.}$ involves only $\sigma$ and $\sigma^\dagger$. Since $[\sigma,\sigma^\dagger]=0$, $\sigma$ commutes with $H_{\rm EK}$ and remains free. In contrast $[\sigma',H_{\rm EK}] \ne 0$, the second parafermion $\sigma'$ develops a non-trivial scaling in the infrared as determined below.
It is more convenient to change to a charge basis that simultaneously diagonalizes $\sigma$ and $\sigma^\dagger$. The rotated clock operators take the new expressions
\begin{equation}\label{rotated-mat}
     \sigma=\left(
\begin{array}{ccc}
 1 & 0 & 0 \\
 0 & \omega & 0 \\
 0 & 0 & \omega^2 \\
\end{array}
\right) \qquad \qquad 
\tau=\left(
\begin{array}{ccc}
 0 & 0 & 1 \\
 1 & 0 & 0 \\
 0 & 1 & 0 \\
\end{array}
\right)
\qquad \qquad  \sigma'=\left(
\begin{array}{ccc}
 0 & 0 & 1 \\
 \omega & 0 & 1 \\
 0 & \omega^2 & 0 \\
\end{array}
\right) 
\end{equation}
and the Hamiltonian turns into a block diagonal form
\begin{equation}\label{bb}
    H_{\rm EK} = 2 J \left(
\begin{array}{ccc}
 \cos\left(\sqrt{\frac{2}{3}}\delta\phi_A\right) & 0 & 0 \\
 0 & \cos\left(\sqrt{\frac{2}{3}}\delta\phi_A+\frac{2\pi}{3}\right) & 0 \\
 0 & 0 & \cos\left(\sqrt{\frac{2}{3}}\delta\phi_A-\frac{2\pi}{3}\right) \\
\end{array}
\right)= H^0 \oplus H^+ \oplus H^-
\end{equation}
The three blocks correspond to the charge sectors $r=0,\pm 1$.
In this rotated basis of the original charge states, we find three decoupled boundary sine-Gordon models, related to each other by just a $\mathbb{Z}_3$ circular shift of the field $\delta \phi_A \to \delta \phi_A + 2\pi/\sqrt{6}$. They all have the same Bethe-ansatz solution describing the crossover from high energies to the infrared, the same in fact as an impurity~\cite{Kane_1992} in a one-dimensional electron gas with Luttinger parameter $K=1/3$. In particular, the residual entropy is predicted~\cite{fendley1994} to decrease by $\Delta S = \ln (\sqrt{3}) = \frac{\ln 3}{2}$ along the crossover. Starting with $S = \ln (3)$ at high energy where the three charge impurity states decouple from the leads, it predicts the fractional entropy $S = \ln (3)-\Delta S = \frac 1 2 \ln (3)$ at the infrared fixed point, as confirmed by NRG calculations (see Fig. 2 in the main text and Ref.~\cite{pouse2023}). The same prediction was moreover obtained~\cite{KBM} at large transparency using a different approach. 

Overall, the physical picture at the stable fixed point is that the parafermion $\sigma'$ is fully screened in the infrared by the boundary term Eq.~\eqref{bb} and one is left with a single free parafermion $\sigma$. The original charge space is thus partially screened dividing the original entropy $\ln (3)$ by a factor $2$.

\subsection{Scaling dimension of $\tau$ at the low-energy fixed point}\label{scalingtau}

The scaling exponent of the screened parafermion $\tau$ can be obtained at low energy without resorting to the Bethe ansatz exact solution describing the RG flow. The picture in the infrared is that the boson field $\delta \phi_A$ is pinned to a different value 
$\delta \phi_A = 0,\mp \sqrt{2/3} \, \pi$ in each charge sector $r=0,\pm 1$. As $\tau$ switches between the (rotated) charge sectors, see Eq.~\eqref{rotated-mat}, it modifies abruptly the scattering for the bosons $\delta \phi_{\rm A} (x)$ controlled by boundary term Eq.~\eqref{bb}. The corresponding orthogonality catastrophe~\cite{FG1} results in a non-trivial dimension for $\tau$ as we will now evaluate.

We are interested  in the time correlator
\begin{equation}\label{time-correlator}
    \langle \tau^\dagger (t) \tau (0) \rangle = \langle e^{i H t/\hbar} \tau^\dagger e^{-i H t/\hbar} \tau \rangle
\end{equation}
where $H = H_0 + H_{\rm EK}$ and the average is taken in the ground state.  $| {\rm GS}_r, r\rangle$ denotes the ground state of $H$ in the sector $r=0,\pm 1$, whereas the second $r$ index indicates the charge sector of the wavefunction. Since $H$ acts diagonally in the charge sectors, we have the obvious identity $H_{\rm EK} | {\rm GS}_r, r\rangle = H^r | {\rm GS}_r, r\rangle$ and, for instance,
\begin{equation}\nonumber
H_{\rm EK} \left( \tau | {\rm GS}_{-}, -\rangle \right) = H_{\rm EK} | {\rm GS}_{-}, 0\rangle
= H^0 \tau | {\rm GS}_{-}, -\rangle,
\end{equation}
where we have used the fact that $\tau$ rotates from $-$ to $0$. With these rules in mind, we average Eq.~\eqref{time-correlator} over the state $| {\rm GS}_{-}, -\rangle$ and rewrite
\begin{equation}\label{corre1}
     \langle {\rm GS}_{-}, - | e^{i H t/\hbar} \tau^\dagger e^{-i H t/\hbar} \tau | {\rm GS}_{-}, -\rangle =  \langle {\rm GS}_{-}, - | e^{i H^{-} t/\hbar}  e^{-i H^0 t/\hbar} | {\rm GS}_{-}, -\rangle.
\end{equation}
We make progress by defining the unitary operator
\begin{equation}\label{vertop}
P \equiv e^{i\sqrt{\frac{2}{3}}\;\Theta_{\rm A}}
\end{equation}
with $\Theta_{\rm A}$ being the conjugate field to $\delta\phi_{\rm A}$ satisfying $\Big[\delta\phi_{\rm A}, \Theta_{\rm A} \Big]=i\pi$. $P$ is a translation operator for the variable $\delta\phi_{\rm A}$, namely \begin{equation}
P \left(  \sqrt{\frac{2}{3}}\delta\phi_A \right) P^{-1} = \sqrt{\frac{2}{3}}\delta\phi_A + \frac{2 \pi}{3},
\end{equation}
permuting circularly the Hamiltonians $P H^r P^{-1} = H^{r+1}$. $P$ is moreover transparent for the kinetic part $H_0$. With these identities, we can rewrite Eq.~\eqref{corre1} as
\begin{equation}
   \langle {\rm GS}_{-}, - | e^{i H^{-} t/\hbar} P e^{-i H^{-} t/\hbar} P^{-1} | {\rm GS}_{-}, -\rangle =
\langle    {\rm GS}_{-}, - | P(t) P^{-1} (0) | {\rm GS}_{-}, -\rangle .
\end{equation}
The average can also be performed over the states $| {\rm GS}_{0}, 0\rangle$, $| {\rm GS}_{+}, +\rangle$ with the same outcome such that we eventually prove the identity
\begin{equation}
 \langle \tau^\dagger (t) \tau (0) \rangle = \langle P (t) P^{-1} (0) \rangle,
\end{equation}
and we can identify  
\begin{equation}\label{identification}
\tau = P^{-1} = e^{-i\sqrt{\frac{2}{3}}\;\Theta_{\rm A}}
\end{equation}
at the infrared fixed point (up to an unknown phase factor). Since the variable $\delta \phi_{\rm A}$ is essentially pinned at low energy, it implies that the conjugate field  $\Theta_{\rm A}$ is free with the zero-temperature correlation function $\langle \Theta_{\rm A} (t) \Theta_{\rm A} (0) \rangle \sim - \ln t$. We find the power law
\begin{equation}
\langle \tau^\dagger (t) \tau (0) \rangle \sim \frac{1}{t^{2/3}},
\end{equation}
corresponding to a scaling dimension $1/3$ for the operator $\tau$. The operator $\sigma'= \sigma \tau$ exhibits the same scaling exponent.

It is straightforward to verify that the operator $\eta = \tau \, e^{i\sqrt{\frac{2}{3}}\;\Theta_{\rm A}}$ commutes with the Hamiltonian~\eqref{bb}, and not only at low energy, and thus defines a constant of motion. This is consistent with the identification $\tau = P^{-1}$.

\section{Scaling exponents of the conductance}

\subsection{Relevant and irrelevant perturbations}\label{Sec:VA}

Despite being a deformation of the original (double-charge) Kondo model (irrelevant in the renormalization group sense), the Toulouse Hamiltonian captures the exponents characterizing the vicinity of the quantum critical point. Moving away from the triple point in the stability diagram of gate voltages is described by $h_{L/R} \ne 0$. $h_L$ and $h_R$ couple to the charge pseudospin operators, written in the original basis $|N,M\rangle/|N+1,M\rangle/|N,M+1\rangle$ as $S_L^z = (1/3) \, {\rm diag} (-1,2,-1)$ and  $S_R^z = (1/3) \, {\rm diag} (2,-1,-1)$ up to unimportant constant terms. They can also be expressed as
\begin{equation}\label{spinoperators}
S_L^z = \frac{1}{3} \left( \omega \tau^\dagger + \omega^* \tau \right), \qquad \qquad 
S_R^z = \frac{1}{3} \left( \omega^* \tau^\dagger +  \omega \tau \right),
\end{equation}
and thereby inherit the scaling dimension $\Delta_R = 1/3$ as $\tau$ in the infrared.


The remarkable property that the boundary term $H_{\rm EK}$ depends on $\sigma$ and $\sigma^\dagger$ which commute with each other holds only for perfect channel symmetry $J_{\rm L}=J_{\rm R}=J_{\rm C}=J$. We describe a weak channel asymmetry with $J_{\rm L}=J_{\rm R}=J-\delta J$ and $J_{\rm C}=J+2 \delta J$ and write the Hamiltonian as $H_{\rm EK} =  H_{\rm EK} (\delta J=0) + 2 \delta J \, \mathbb{Q}$ with
\begin{equation}
    \mathbb{Q} =  \tau \,  \left(
\begin{array}{ccc}
 \cos\left(\sqrt{\frac{2}{3}}\delta\phi_A+\frac{2\pi}{3}\right) & 0 & 0 \\
 0 & \cos\left(\sqrt{\frac{2}{3}}\delta\phi_A-\frac{2\pi}{3}\right) & 0 \\
 0 & 0 & \cos\left(\sqrt{\frac{2}{3}}\delta\phi_A\right) \\
\end{array}
\right) + h.c.
\end{equation}
in the rotated basis.
Following the same steps as Sec.~\ref{scalingtau}, we find that $\mathbb{Q}$ obeys the same scaling dimension $\Delta_R = 1/3$ as $\tau$, namely
\begin{equation}
\langle \mathbb{Q}^\dagger (t) \mathbb{Q} (0) \rangle \sim  \frac{1}{t^{2/3}}    
\end{equation}
The operators $S_L^z$, $S_R^z$ and $\mathbb{Q}$ all destabilize the QCP with the same exponent towards a FL regime where the remaining parafermion is eventually screened (see Fig. 2c in the main text). They generate a FL temperature scale $T^* \sim h_{R}^{1/(1-\Delta_R)}$ (or $h_{L}^{1/(1-\Delta_R)}$, $\delta J^{1/(1-\Delta_R)}$).

So far, we have restricted our investigation to the fine-tuned Toulouse line $\bar{\lambda}=0$. At low temperature, $\bar{\lambda} \ne 0$ governs the leading irrelevant temperature correction to observables. From Eq. (9) in the main text, it involves the operator
\begin{equation}\label{LIO}
    {\cal O}_{\rm LIO} = S^z_{\rm L}
    \partial_x \delta\phi_1(0)+ S^z_{\rm R} \partial_x \delta\phi_2 (0)
\end{equation}
with dimension $4/3 = 1+1/3$, the sum of the dimension $\Delta_R = 1/3$ of $S_{L/R}^z$ ($\tau$) and the dimension $1$ of the operators $\partial_x \delta\phi_{1/2}(0)$, or
\begin{equation}\label{scalingLIO}
\langle {\cal O}_{\rm LIO} (t) {\cal O}_{\rm LIO} (0) \rangle \sim 1/t^{8/3}
\end{equation}
As shown below in Sec.~\ref{linear-conduc}, this operator yields the $(T/T_K)^{2/3}$ low-temperature correction to the conductance which confirms the NRG asymptotics of Fig. 2(b) (main text).

\subsection{Linear conductance}\label{linear-conduc}

The charge current through the double-charge setup can be expressed as
\begin{equation}\label{currentexpression}
    \hat{I} =   - \frac{e}{2 \pi} \sqrt{\frac{2}{3}} \, \partial_t  \Theta_{\rm A} 
\end{equation}
In principle, the full expression also includes the fields $\Theta_{\rm B/D}$ but they carry no average current since the corresponding conjugate fields are free. 
The linear conductance is expressed using the Kubo formula as
\begin{equation}\label{TA}
G=\frac{e^2}{3h}\frac{\omega_n}{\pi}\Big<\Theta_{\rm A} (i\omega_n)\Theta_{\rm A} (-i\omega_n)\Big>_{i\omega_n\to 0^+},
\end{equation}
where $\omega_n=2\pi n T/\hbar$ are bosonic Matsubara frequencies and the Fourier transform is defined as
\begin{equation}
\Theta_{\rm A} (i\omega_n)=\int^{\hbar/T}_0 d\tau e^{i\omega_n\tau} \, \Theta_{\rm A}(\tau).
\end{equation}
In the path integral formulation, the action at the quantum critical point is simply quadratic in $\Theta_{\rm A}$
\begin{equation}
\mathcal{S}_0= \sum_n \frac{|\omega_n|}{2 \pi} \left|\Theta_{\rm A} (i\omega_n)\right|^2.
\end{equation}
The linear conductance at the QCP is readily obtained from the Kubo formula Eq.~\eqref{TA} with the result $G = G_0 = e^2/(3 h)$.
Away from the QCP, the action acquires corrections to the action
\begin{align}
\mathcal{S}_1=&   \int^{\hbar/T}_0 d\tau\;\left[ \mathcal{L}_L (\tau) +\mathcal{L}_R (\tau) \right],
\end{align}
with 
\begin{equation}\label{action-field}
\mathcal{L}_L = - \frac{4 h_L}{3} \cos \left(    \sqrt{\frac{2}{3}}\;\Theta_{\rm A} + \frac{2 \pi}{3} \right) \qquad \qquad
    \mathcal{L}_R  = - \frac{4 h_R}{3} \cos \left(    \sqrt{\frac{2}{3}}\;\Theta_{\rm A} - \frac{2 \pi}{3} \right)
\end{equation}
where we used the operator identification~\eqref{identification}. In addition, energies are cut off at the Kondo temperature scale $T_K$.
The first order correction due to $\mathcal{L}_{L/R}$ vanishes. Expanding the action to second order and evaluating the gaussian integrals~\cite{Furusaki1995b,KBM}, we arrive at
\begin{equation}\label{conduc-correction}
G = \frac{e^2}{3h} \Big[ 1-\mathcal{C}_1 \frac{h_L^2+h_R^2 - h_L h_R}{T_K^2} \left( \frac{T_K}{T} \right)^{4/3}  \Big],
\end{equation}
where $C_1$ is a dimensionless coefficient which can be absorbed into a redefinition of the Kondo temperature. The combination $h_R^2+h_L^2-h_L h_R$ predicts an anisotropic conductance in the plane of gate voltages (stability diagram). Rephrased with the energies of the individual charge states, $\delta E_{1/2/3}$, measured relative to the QCP, it takes the symmetric form $h_R^2+h_L^2-h_L h_R \sim \delta E_1^2 + \delta E_2^2 +\delta E_3^2$.

A finite $h_L$ or $h_R$ destabilizes the QCP with unitary conductance $G_0$ and drives the system towards a new Fermi liquid point with zero entropy and vanishing conductance. The exact same crossover is driven by channel asymmetry $\delta J \ne 0$. The onset of this relevant perturbation is given by Eq.~\eqref{conduc-correction}, with the behaviour $(T/T^*)^{-4/3}$. The crossover scale $T^*$ depends on the pseudo-magnetic fields as $T^* \sim (h_L^2+h_R^2 - h_L h_R)^{3/4}/\sqrt{T_K}$ (or $T^* \sim \delta J^{3/2}/\sqrt{T_K}$), or $T^* \sim N_g^{3/2}$ where $N_g$ is the gate voltage distance to the triple point, in agreement with our NRG data and the results of Ref.~\cite{pouse2023,KBM}. 

The finite temperature correction at the QCP can be similarly evaluated by including the leading irrelevant operator  $\bar{\lambda} {\cal O}_{\rm LIO}$ (Eq. (9), main text). The action is supplemented by
\begin{equation}
\mathcal{S}_{\rm LIO} = \bar{\lambda} \int^{\hbar/T}_0 d\tau\;  {\cal O}_{\rm LIO} (\tau),
\end{equation}
where the imaginary-time correlator of the leading irrelevant operator is deduced from Eq.~\eqref{scalingLIO} by conformal invariance
\begin{equation}
    \langle {\cal O}_{\rm LIO} (\tau) {\cal O}_{\rm LIO} (0) \rangle = \frac{1}{T_K^{2/3}} \left( \frac{\pi T}{T_K \sin ( \pi T \tau)}  \right)^{8/3} 
\end{equation}
The linear conductance is finally computed perturbatively to second order in $\bar{\lambda}$
\begin{equation}
G = \frac{e^2}{3h} \left[ 1-\mathcal{C}_{LIO}  \left( \frac{T}{T_K} \right)^{2/3}  \right],
\end{equation}
where $\mathcal{C}_{LIO} \propto \bar{\lambda}^2$ is a dimensionless prefactor. The first order in $\bar{\lambda}$ vanishes. The exponent agrees with the experimental results of Ref.~\cite{pouse2023} as well as with the power law extracted from the NRG (Fig. 2(a) in the main text) and the quasi-ballistic power law in Ref.~\cite{KBM}.


\subsection{Non-linear current}

The effect of a finite voltage biasing of the two-charge Kondo circuit is readily addressed close to the triple point (QCP). The field $\Theta_{\rm A}$, conjugate to $\delta\phi_{\rm A}$, is a sum of chiral fields
\begin{equation}
\Theta_{\rm A} = \frac{\Theta_{\rm A,R}-\Theta_{\rm A,L}}{\sqrt{2}}
\end{equation}
moving in opposite directions. The outgoing field $\Theta_{\rm A,R}$ acquires a time dependence with the applied voltage $V$
\begin{equation}
    \Theta_{\rm A,R} \to \Theta_{\rm A,R} - \frac{1}{\sqrt{3}} \frac{e V t}{\hbar} 
\end{equation}
whereas the incoming field $\Theta_{\rm A,L}$ is unchanged.
Right at the triple point, inserting this time dependence into the current expression Eq.~\eqref{currentexpression} directly recovers the unitary form $I = G_0 V$ where $G_0 = e^2/(3 h)$. In the  interaction representation, the Hamiltonian corresponding to Eq.~\eqref{action-field} at finite $h_{R/L} \ne 0$ takes the form
$H_{R/L} (t) = {\cal T}_{R/L} (t) + {\cal T}_{R/L}^\dagger (t)$, with the operator 
\begin{equation}
    {\cal T}_{R/L} (t) = - \frac{2 h_{R/L}}{3} \, \exp \left (i   \sqrt{\frac{2}{3}}\;\Theta_{\rm A} \mp \frac{2 \pi}{3}  -i \frac{e V t}{3} \right).
\end{equation}
Then, following linear response theory, the current operator Eq.~\eqref{currentexpression} is expanded in powers of $H_{R/L}$. A careful analysis must account for the fact that the current operator is taken at a position $x=\ell$ distant from the outermost right QPC located at $x=0$: ${\cal T}_{R/L}  \equiv {\cal T}_{R/L} (0)$ in $H_R$ and $\hat{I} \equiv \hat{I} (\ell)$ in Eq.~\eqref{currentexpression}. Using the commutation relations
$$
[ \partial_t \Theta_{\rm A,R/L} (t,\ell) , \Theta_{\rm A,R/L} (t',0) ] = - 2 i \pi \delta ( t-t' \mp \ell/v_F) 
$$
expressing causality, we obtain the expansion
$\hat{I} = \hat{I}_0 + \hat{I}_1+\hat{I}_2$ with ~\cite{morel2021,KBM}
\begin{equation}\label{currents}
\begin{split}
\hat{I}_0 (t) &= - \frac{e}{2 \pi} \sqrt{\frac{2}{3}} \, \partial_t  \Theta_{\rm A} (t,\ell) + \frac{e^2 V}{3 h} \\
\hat{I}_1 (t) &= i \, \frac{e^*}{\hbar} \sum_{j=R/L} [{\cal T}_j(t - \ell/v_F)-{\cal T}_j^\dagger(t-\ell/v_F)]   \\
\hat{I}_2 (t) &= \frac{e^*}{\hbar^2}  \int_{-\infty}^{t_r}\text{d}t'\,  \sum_{j,j'=R/L} [{\cal T}_j (t') + {\cal T}_j^\dagger(t'),{\cal T}_{j'}^\dagger (t_r) - {\cal T}_{j'} (t_r)]
    \end{split}
\end{equation}
where we introduced the fractional charge $e^*=e/3$ and $t_r = t-\ell v_F$ accounting for the transit time from the right QPC. The first order term $\hat{I}_1$ has a vanishing average whereas~\cite{morel2021,KBM}
\begin{equation}
\langle \hat{I}_2 \rangle = - e^* \frac{2 \pi}{\Gamma(2/3)} \left( \frac{2 }{3 \hbar} \right)^2    (h_L^2+h_R^2-h_L h_R) \left( \frac{3 \hbar}{e V} \right)^{1/3} t_0^{2/3}.
\end{equation}
$t_0 \sim \hbar/T_K$ is the short-time cutoff of the effective model. The final result for the total current is thus
\begin{equation}
I = \frac{e^2 V}{3h} \left[ 1-\mathcal{D}_1 \frac{h_L^2+h_R^2-h_L h_R}{T_K^2} \left( \frac{T_K}{e V} \right)^{4/3}  \right],
\end{equation}
with the dimensionless coefficient $\mathcal{D}_1$, or a conductance correction $\sim (V/T^*)^{-4/3}$. A similar calculation shows that the channel asymmetry $\delta J$ yields exactly the same scaling $\sim (V/T^*)^{-4/3}$ with $T^* \sim \delta J^{3/2}/\sqrt{T_K}$.

We can also utilize 
Eq.~\eqref{currents} to predict the shot noise. Following similar calculations in Refs.~\cite{morel2021,KBM}, we obtain the Fano factor 
\begin{equation}
F= \frac{S}{I(h_{R/L}=0)-I}= \frac 1 3
\end{equation}
corresponding to the backscattering of fractional charges $e^*=e/3$. The very same prediction was done in the  quasi-ballistic limit investigated in Ref.~\cite{KBM} suggesting that the scattering of fractional charges is independent of the transmission of the QPCs and requires only proximity to the QCP. This somewhat extends the prediction of Ref.~\cite{Sela_2018} where a charge $e^*=e/2$ was scattered in the two-channel Kondo model. Here, the $\mathbb{Z}_3$ parafermion scatters the charge $e^*=e/3$ and it should be straightforward to show that cascading $N$ consecutive islands isolates a free $\mathbb{Z}_{N+1}$ parafermion scattering $e^*=e/(N+1)$ charges. This result was confirmed close to the ballistic regime in Ref.~\cite{KBM}.


\section{NRG calculations}
The NRG calculations presented in the main text were performed on the model Eq.~\ref{eq:Hdck2}, in which the two retained charge states of each island in Eq.~\ref{eq:states} are mapped to `impurity' spin-$\tfrac{1}{2}$ degrees of freedom $\hat{\boldsymbol{S}}_L$ and $\hat{\boldsymbol{S}}_R$. Since we are interested in the universal behavior of the critical point arising at a triple point (TP) of the charge stability diagram \cite{pouse2023}, we consider explicitly the limit where states $|A\rangle \equiv |N,M\rangle$, $|B\rangle \equiv |N+1,M\rangle$ and $|C\rangle \equiv |N,M+1\rangle$ all have the same energy $E_A=E_B=E_C\equiv E$ (when $J_L=J_R=J_C=0$), while $|D\rangle \equiv |N+1,M+1\rangle$ has much higher energy $E_D\ggg E$, and is projected out. This is achieved in practice by setting $h_L=h_R=-\tfrac{1}{2}I$ and sending $I\to +\infty$. The influence on the physics of neighbouring TPs is therefore eliminated. This is justified at low temperatures if we focus on the close vicinity of the critical point, since the capacitive interaction $I$ is RG irrelevant. This reveals the universal physics of the critical point most cleanly. We therefore study with NRG the reduced model for the TP,
\begin{eqnarray}
\label{eq:Htp}
H_{\rm TP} =  H_{\rm leads} + \left (J_L^{\phantom{-}} \hat{s}_{L}^- |B\rangle\langle A| + J_C^{\phantom{-}}\hat{s}_{C}^- |C\rangle\langle B| + J_R^{\phantom{-}}\hat{s}_{R}^- |A\rangle\langle C| + {\rm H.c.} \right )  \;,
\end{eqnarray}
which is a highly nontrivial generalized quantum impurity model, featuring a three-state impurity whose configurations are interchanged by scattering between six spinless conduction electron channels (physically, these channels are the electronic reservoirs either side of each of the three QPCs in the two-island device). 

We focus on (i) the critical point arising with finite $J_L=J_R=J_C\equiv J$; and (ii) the Fermi liquid crossover generated by perturbing away from the critical point by setting $J_L=J_R\equiv J$ but $J_C \ne J$.

The model is then solved using a variant of Wilson's NRG method \cite{wilson1975renormalization,*bulla2008numerical}, in which the channels are interleaved in a generalized Wilson chain \cite{mitchell2014generalized,*stadler2016interleaved}, coupled at one end to the impurity. This more efficient `iNRG' method is required for a model of such high complexity, especially along the experimentally-relevant Fermi liquid crossover where symmetries are broken.

For all iNRG calculations presented in this work, we use a logarithmic discretization parameter $\Lambda=4$, retain $N_s=35000$ states at each iteration, and exploit all abelian quantum numbers. The impurity contribution to the entropy $S_{\rm imp}(T) = S_{\rm tot}(T)-S_{\rm elec}(T)$ is obtained in the usual way for NRG from the partition function \cite{wilson1975renormalization,*bulla2008numerical}, where $S_{\rm tot}$ is the full entropy for the coupled impurity-lead system, while $S_{\rm elec}$ is the entropy of the isolated free electronic reservoirs.

The series dc linear response differential conductance,
\begin{equation}\label{eq:cond}
G=\frac{dI}{dV}\Bigg|_{V\to 0}
\end{equation}
is defined in terms of the current $I=-e\langle \dot{N}_{R\uparrow}\rangle$ flowing into the drain lead ($R\uparrow$ in Fig.~1) due to a bias voltage $V$ applied to the source lead ($L\uparrow$). Here $\dot{N}_{R\uparrow}=\tfrac{d}{dt}\hat{N}_{R\uparrow}$ and $\hat{N}_{\alpha \uparrow}=\sum_{k}\psi_{\alpha \uparrow k}^{\dagger}\psi_{\alpha \uparrow k}^{\phantom{\dagger}}$. An ac voltage bias on the left lead can be incorporated by a source term in the Hamiltonian, $H_{\rm bias}=-eV\cos(\omega t) \hat{N}_{L \uparrow}$, where $\omega$ is the ac driving frequency. The dc limit is obtained as $\omega \to 0$. 

We use the Kubo formula \cite{Galpin2014} to obtain the desired conductance,
\begin{equation}\label{eq:kubo}
    G = \frac{e^2}{h}\lim_{\omega \to 0} \frac{-2\pi  ~{\rm Im}K(\omega)}{ \omega} \;,
\end{equation}
where $K(\omega)=\langle\langle \dot{N}_{L \uparrow} ; \dot{N}_{R \uparrow}\rangle\rangle$ is the Fourier transform of the \emph{equilibrium} retarded current-current correlator\\ $K(t)=-i\theta(t)\langle [\dot{N}_{L \uparrow },\dot{N}_{R \uparrow}(t) ] \rangle$. Within iNRG, ${\rm Im}K(\omega)$ may be obtained from its Lehmann representation using the full density matrix technique \cite{weichselbaum2007sum} in terms of the Anders-Schiller basis \cite{anders2006spin} established on the iNRG generalized Wilson chain \cite{mitchell2014generalized,*stadler2016interleaved}. The numerical evaluation is substantially improved by utilizing the identity ${\rm Im}K(\omega)=\omega^2 {\rm Im}\langle\langle \hat{N}_{L \uparrow} ; \hat{N}_{R \uparrow}\rangle\rangle$ as shown recently in Ref.~\cite{transport}. 
We use this method to obtain the NRG conductance results presented in the main paper.


\section{Experimental data fitting}
The experimental data of Fig.~3 were taken from Ref.~\cite{pouse2023}, where a full description of the experimental setup is given. For a particular configuration of $\tau$ and $\tau_C$, conductance is measured while the voltages on two gates, one coupled to each island, are varied. Then, line cuts (inset of Fig.~3) are extracted along the line between a pair of TPs with equal left and right island gate voltage detuning $U$, where $U = 0$ is defined as the point midway between a pair of TPs. For a given setting of $\tau and \tau_C$, multiple line cuts (across different pairs of TPs in the same 2D plot, or the same pair of TPs in successively acquired 2D plots) are averaged to reduce the noise. In measurements for $\tau_C = 0.998$ (red), a charge near the islands evidently switched to a new location and then switched back, producing an apparent discontinuity in the curve (see Supplementary Fig. 6 of Ref.~\cite{pouse2023} for an example.) However, this does not contribute to the scaling collapse analysis, which only uses the tail of the line cut, not the portion between the triple points. In this case we specifically used the righthand tail (positive $U$), though using the lefthand tail (or both tails) would not have changed the results in any substantive way.

$T^*$ is determined using the expression $T^* = T_0^* + b |\cos{(2 \pi U/\delta_{TP})} - \Delta_{TP} |^{3/2}$~\cite{KBM}, which accounts for the periodic structure of the TPs. $\delta_{TP}$ corresponds to the spacing from one TP pair to another TP pair, and $\Delta_{TP}$ to the splitting between the TPs of a single pair. Both are experimentally determined for each $\tau_C$ line cut from longer line cuts spanning multiple pairs of TPs from the same stability diagram or other charge stability diagrams with the same $\tau, \tau_C$ values. The prefactor $b=1$~mK is the same value used in Ref.~\cite{pouse2023}, which is found from fitting $b$ of each line cut to best match an NRG calculated universal curve and then averaging the resulting $b$ values of all four line cuts. $T_0^*$ accounts for the detuning in $\tau_C$ and is determined for each line cut by a fit, such that the line cut best collapses onto the others. To fix an overall free parameter in the definition of the Fermi liquid scale $T^*$, we take a practical definition in which it corresponds to the conductance half-width-at-half-maximum at the TP, such that $G(T=T^*)=G_0/2$, with $G_0=e^2/3h$ the critical point conductance.

From the analytic form for the leading corrections to the critical conductance given in Eq.~13 of the main text, $c_K$, $T_K$ and $c^*$ are left as free parameters in a fit to the experimental data. However, since the value of $\tau=0.95$ is kept fixed for all data sets shown (it is $\tau_C$ and hence $T^*$ that is varied), the value of $T_K$ is taken to be a constant. Therefore the correction to the fixed point conductance $G_0=e^2/3h$ at $T^*=0$ is taken to be a constant, $G_0 - G(T^*=0) \simeq 0.021$ $e^2/h$. This yields $T_K / (c_K)^{3/2} = 6.42$ K (we do not separately determine the prefactor $c_K$ since its value is somewhat arbitrary, depending on the definition of $T_K$ used). The remaining variation in the conductance as a function of $\tau_C$  and $|U|$ in the data plotted in Fig.~3 is captured by the $T^*$ term in Eq.~13, from which our fit to experimental data yields $c^* \simeq 0.23$. The fit is also consistent with our NRG results.


\section{Parafermions in Kondo models}

The emergence of a free (unscreened) Majorana fermion in the two-channel Kondo model has been established with an explicit construction via the Emery-Kivelson method~\cite{EK1}. 
It has also been reproduced by starting from the opposite quasi-ballistic limit related to the two-channel charge Kondo model~\cite{Furusaki1995b}. Some other Kondo models have been argued to host free parafermions at their fixed point. However the identification was done by analogy of the residual entropy with the quantum dimension associated with a parafermion operator, not with an explicit derivation, in contrast with what is done in the main text of this Letter and in Sec.~\ref{seciv}, where we construct the parafermion operator and relate it to the original variables of the model. The ability to do this had previously been hindered because models hosting parafermions are irreducibly interacting, and therefore more challenging to solve than their non-interacting counterparts that can host the more simple Majorana modes.

Let us review which Kondo models exhibit a residual entropy suggestive of a local unscreened parafermion mode. The standard multi-channel Kondo model with $N$ channels has the residual entropy~\cite{AFFLECK_AP_1995}
\begin{equation}
    S = \ln \left[ 2 \cos \frac{\pi}{2+N} \right].
\end{equation}
The result $S = \ln [(1+\sqrt{5}/2)]$ in the three-channel case suggests a local Fibonacci anyon~\cite{Pierre_2018}. In the four-channel Kondo model~\cite{FG1}, which has similarities with the model studied here, $S = \frac{1}{2} \ln (3)$ indicating a local $\mathbb{Z}_3$ parafermion.
The topological Kondo model with $SO(M)$ symmetry~\cite{beri2017,*altland2014bethe}, the residual entropy is $S= \frac{1}{2}\ln M$ for $M$ odd and $S= \frac{1}{2}\ln (M/2)$ for $M$ even, suggesting a local parafermion in all cases and notably a $\mathbb{Z}_3$ parafermion for $M=3$.

$\mathbb{Z}_3$ parafermions have also been mentioned in the context of the three-channel charge Kondo model~\cite{Pierre_2018} but with a totally different meaning. There, the renormalization group flow starting from the quasi-ballistic limit has been argued~\cite{affleck2001} to map onto the boundary three-state Potts model. The $\mathbb{Z}_3$ parafermions are then bulk operators that appear in the conformal field theory description of the Potts model. Firstly, those $\mathbb{Z}_3$ parafermions are delocalized objects extending in the leads and differ from the local operators discussed in this work. Secondly, the fixed point of the three-channel Kondo model does not seem to decouple one of the $\mathbb{Z}_3$ parafermions but rather a Fibonacci anyon as indicated by the residual entropy.


\section{Kondo vs quasi-ballistic limits}

The physical device studied experimentally in Ref.~\cite{pouse2023} consists of two hybrid metal-semiconductor islands, both of which host a macroscopically-large number of charge states. The islands have a finite capacitance and hence a finite charging energy $E_C$, which in practice is found to be much larger than the experimental base temperature, $k_B T$. The islands are connected to each other and to metallic leads by QPCs. The island-lead transmission $\tau$ and the island-island transmission $\tau_C$ can be tuned \textit{in-situ} within a single device from the weak-tunneling (Kondo) limit ($\tau, \tau_C \ll 1$) through to the quasi-ballistic limit ($\tau, \tau_C \sim 1$). Ref.~\cite{pouse2023} studied both limits, and the evolution of behavior between them, experimentally. 

However, no single theoretical technique can treat this system exactly for all values of the QPC transmissions. In this Letter we have considered the weak-tunneling (Kondo) limit using our modified EK approach \cite{EK1} and NRG \cite{wilson1975renormalization,*bulla2008numerical}. In this limit, one can rigorously derive \cite{pouse2023} a low-energy effective model in which only two charge states per island are retained -- the DCK model. The EK and NRG methods employed here are suited to analyzing such generalized quantum impurity models. The regime of applicability of NRG was extended to intermediate transmission in Ref.~\cite{pouse2023} by generalizing the DCK model to include several (but still a finite number) of charge states per island. However, larger transmission remains out of reach for EK and NRG because a diverging number of charge states become involved in transport \cite{Furusaki1995b} and the quantum impurity model description breaks down. On the other hand, the quasi-ballistic limit, near perfect transmission, was studied recently in Ref.~\cite{KBM} using Matveev's bosonization approach \cite{Furusaki1995b}. In that work, a continuum version of the model is mapped to almost-free bosons, in which the interaction is treated perturbatively. Different models and methods are therefore used in the two limits. Note however that electronic interactions play a decisive role in both cases.

Remarkably, the low-temperature physics near the critical triple point is the same in both limits: both the DCK model and the quasi-ballistic model capture the same behavior. Fundamentally this is a consequence of universality near the critical point: the rescaled low-energy physics is the same, independently of the microscopic details of the bare model, such as the values of the bare transmissions. This universality can be exploited to use results obtained at small transmission and apply them in the experimental setting at larger transmission, provided we are at low temperatures and confine attention to the vicinity of the critical point. This correspondence was demonstrated in Ref.~\cite{pouse2023} and again in the present work when comparing theoretical predictions to experimental data. We note that the same approach has been adopted previously for single-island devices connected to two and three channels in e.g.~Refs.~\cite{Pierre_2015,Sela_2016,Pierre_2018}, to establish quantitative agreement between NRG results obtained in the weak-tunneling Kondo regime, and experimental data obtained at larger transmission. The latter is preferable experimentally since then Kondo scales are boosted far above base temperatures, allowing the universal regime to be accessed.


\end{document}